\documentclass{moriond}

\usepackage[utf8]{inputenc}
\usepackage[normalem]{ulem}
\usepackage[english]{babel}%,frenchb
\usepackage{tabularx}
\usepackage{array}
\usepackage{graphics}
\usepackage{psfrag}
\usepackage{epsfig}
\usepackage{subfig}
\usepackage{mathtools}
\usepackage{amssymb}
\usepackage{bbold}
\usepackage{setspace}
\usepackage{rotating}
\usepackage{colortbl}
\usepackage{longtable}
\usepackage{slashed}
\usepackage{braket}
\usepackage{lineno}
\makeatletter
\usepackage{textcomp}
\usepackage[usenames,dvipsnames]{xcolor}
\usepackage{relsize}

\usepackage{float}

\usepackage{bbm}
\usepackage{bm}
\usepackage{enumerate}
\usepackage{amsmath}
\usepackage{pifont}

\def\be{\begin{equation}}
\def\ee{\end{equation}}
\def\bea{\begin{eqnarray}}
\def\eea{\end{eqnarray}}

\newcommand{\xmark}{\ding{55}}%

\newcommand{\gev}{{\rm GeV}}

\newcommand{\pdir}{p\kern -5.2pt\raise 0.2ex\hbox {/}}

\newcommand{\vdir}{v\kern -5.75pt\raise 0.15ex\hbox {/}}
\newcommand{\kdir}{k\kern -5.75pt\raise 0.15ex\hbox {/}}
\newcommand{\epsdir}{\epsilon\kern -5.0pt\raise 0.15ex\hbox {/}}
\newcommand{\bvdir}{\bar{v}\kern -5.75pt\raise 0.15ex\hbox {/}}
\newcommand{\Ddir}{D\kern -7.75pt\raise 0.20ex\hbox {/}}
\newcommand{\Adir}{A\kern -7.75pt\raise 0.20ex\hbox {/}}
\newcommand{\ldir}{l\kern -5.0pt\raise 0.2ex\hbox{/}}

\makeatother

\definecolor{niceblue}{rgb}{0.15,0.15,0.6}
\definecolor{nicegreen}{rgb}{0.1,0.5,0.1}
\definecolor{Red}{rgb}{1.,0.,0.}

\definecolor{Green}{rgb}{0.2,.7,0.2}

\newcommand{\Photo}{}

\begin{document}
\vspace*{4cm}
\title{Single--Leptoquark Solutions to the $B$-physics Anomalies }

\author{Andrei Angelescu}

\address{Department of Physics and Astronomy, University of Nebraska-Lincoln,\\
Lincoln, NE, 68588, USA.}

\maketitle\abstracts{We examine various scenarios in which the Standard Model is extended by a light leptoquark state to explain one or both $B$--physics anomalies. Combining low--energy constraints and direct searches at the LHC, we confirm that the only single leptoquark model that can explain both anomalies at the same time is a vector leptoquark, known as $U_1$. Focusing on $U_1$, we highlight the complementarity between LHC and low--energy constraints, and argue that improving the experimental bound on $\mathcal{B}(B\to K\mu\tau)$ by two orders of magnitude could compromise its viability as a solution to the $B$-physics anomalies.}

\section{Introduction}

Over the past several years, there has been a growing interest in the theoretical studies of the origin of lepton flavor universality violation (LFUV), motivated by a number of experimental hints in weak decays of $B$-mesons pointing towards LFUV. The first such indication concerns the measurement of the ratios
%%%%%%%%%%%%%%%%
\begin{equation}
R_{D^{(\ast)}} = \left. \dfrac{\mathcal{B}(B\to D^{(\ast)} \tau\bar{\nu})}{\mathcal{B}(B\to D^{(\ast)} l \bar{\nu})}\right|_{l\in \{e,\mu\}},
\label{eq:RD_definition}
\end{equation}
%%%%%%%%%%%%%%%%
for which several experimental collaborations found an excess of  
$3.8\,\sigma$~\cite{Lees:2012xj,Lees:2013uzd} with respect to the (w.r.t.) Standard Model (SM) prediction.~\cite{Lattice:2015rga,Na:2015kha,Aoki:2016frl,Bigi:2017jbd,Bernlochner:2017jka}

Another indication of the LFUV came from the weak decays mediated by a flavor changing neutral current (FCNC), $b\to sl^+l^-$. LHCb measured 
%%%%%%%%%%%%%%%%
\begin{equation}
R_{K^{(\ast)}}^{[q_1^2, q_2^2]} =  \dfrac{\mathcal{B}'(B\to K^{(\ast)} \mu\mu)}{\mathcal{B}'(B\to K^{(\ast)} ee)}  \,,
\label{eq:RK_definition}
\end{equation}
%%%%%%%%%%%%%%%%
where $\mathcal{B}'$ stands for the partial branching fraction obtained by integrating over $q^2=(p_{l^+}+p_{l^-})^2$  between $q_1^2$ and $q_2^2$ (in units of $\gev^2$), and found a value $\approx 2.5\,\sigma$~\cite{Aaij:2014ora,Aaij:2017vbb} ($4\sigma $ when comined) lower than the value predicted in the SM~\cite{Bordone:2016gaq}. The observations that $R_{D^{(\ast)}}^\mathrm{exp}> R_{D^{(\ast)}}^\mathrm{SM}$ and $R_{K^{(\ast)}}^\mathrm{exp}< R_{K^{(\ast)}}^\mathrm{SM}$ are commonly referred to as the ``{\it $B$-physics anomalies}".

In order to explain these anomalies, physics beyond the SM (BSM) is needed. One of the most popular BSM scenarios for addressing the anomalies involve leptoquarks (LQs), which are colored new bosons that couple to both quarks and leptons. In this work, based on Ref.~\cite{Angelescu:2018tyl}, we study the possibility of a single scalar or vector LQ as a mediator of new physics (NP) that can accommodate one or both of the $B$-physics anomalies, i.e. $R_{K^{(\ast)}}^\mathrm{exp}< R_{K^{(\ast)}}^\mathrm{SM}$ and/or $R_{D^{(\ast)}}^\mathrm{exp}> R_{D^{(\ast)}}^\mathrm{SM}$. In doing so, we combine the updated constraints arising from numerous low-energy physics observables with those deduced from the direct searches at the LHC. Since this kind of models can give rise to lepton flavor violation (LFV),~\cite{Glashow:2014iga} a particular attention is devoted to the vector LQ model $U_1$, for which we will show that the improved experimental bounds on $\mathcal{B}(B\to K^{(\ast )}\mu\tau)$ could validate or discard the model in its minimal form. 

\section{Effective Theory Description}
\label{sec:eft}

In this Section, we recall the low-energy effective theory relevant to both $b\to s\ell^+\ell^-$ and $b\to c\ell \bar \nu$ decays, focusing only on $(V-A)\times(V-A)$ four--fermion operators, which provide a viable explanation for both $b\to s\ell^-\ell^+$ and $b\to c\ell\bar \nu$ $B$--physics anomalies.

\subsection{$b\to s\ell_1^-\ell_2^+$ and $R_{K^{(\ast)}}$}
\label{ssec:RK}

Since we are concerned with both lepton flavor conserving and LFV decay modes, we describe the effective $(V-A)\times(V-A)$ Hamiltonian for a generic $b\to s \ell_1^- \ell_2^+$ transition, with $\ell_{1,2}\in\{e,\mu,\tau\}$, which can be written as

\begin{equation}
\label{eq:hamiltonian-bsll}
  \mathcal{H}_{\mathrm{eff}} \supset -\frac{4
    G_F}{\sqrt{2}}V_{tb}V_{ts}^* \frac{e^2}{(4\pi)^2} \left[ C_{9}^{\ell_1\ell_2} \, (\bar{s}\gamma_\mu P_{L}
    b)(\bar{\ell}_1\gamma^\mu\ell_2) + C_{10}^{\ell_1\ell_2} \, (\bar{s}\gamma_\mu P_{L} b)(\bar{\ell}_1\gamma^\mu\gamma^5\ell_2) \right] +\mathrm{h.c.} \,
\end{equation}
where $C_{9,10}^{\ell_1 \ell_2}$ are the Wilson coefficients relevant to our study. By assuming that the NP couplings to electrons are negligible, it has been established that $R_K$ and $R_{K^\ast}$ can be explained by a purely vector Wilson coefficient, $C_9^{\mu\mu}<0$, or by a left-handed combination, $C_9^{\mu\mu}=-C_{10}^{\mu\mu}<0$, as NP couplings to electrons are disfavored (see e.g. Ref.~\cite{Capdevila:2017bsm}). The result of our fit is

\begin{equation}
\label{eq:C9-exp}
C_9^{\mu\mu}=-C_{10}^{\mu\mu} \in (-0.85,-0.50)\,,
\end{equation}
which deviates from the SM by almost $4\sigma$. In this fit we used  $R_{K^{(\ast)}}^{\mathrm{exp}}$,~\cite{Aaij:2014ora,Aaij:2017vbb} and the theoretically clean $\mathcal{B}(B_s\to \mu\mu)^{\mathrm{exp}}=\left(3.0\pm 0.6^{+0.3}_{-0.2}\right) \times 10^{-9}$.~\cite{Aaij:2017vad,Bobeth:2013uxa} The possibility of having $C_9^{\mu\mu}=-C_{10}^{\mu\mu}$ is particularly interesting because it is realized in several LQ scenarios.~\cite{Dorsner:2016wpm} From now on, for notational simplicity, we will omit the ``$\mu\mu$"-superscript.

\subsection{$b\to c\ell\bar \nu$ and $R_{D^{(\ast)}}$}
\label{ssec:RD}

On the charged current--side, the relevant dimension-six effective Lagrangian reads
\begin{equation}
\label{eq:lagrangian-lep-semilep}
\mathcal{L}_{\mathrm{eff}} \supset -2\sqrt{2}G_F V_{u d} \left[ (1+g_{V_L})\,(\overline{u}_{L}\gamma_\mu {d}_{L})(\overline{\ell}_L\gamma^\mu\nu_L) \right]+\mathrm{h.c.}\,,
\end{equation}
where $u$ and $d$ stand for a generic up- and down-type quarks, and $g_{V_L}$ is the effective NP coupling (which amounts to a rescaling of the SM contribution).

To determine the allowed values of $g_{V_L}$, we assume that NP only contributes to the transition $b\to c\tau \bar{\nu}$, and that its effect is negligible to the electron and muon modes.~\footnote{That assumption is a very good approximation to the realistic situation. As we shall see, we find that the couplings of leptoquarks to $b$ and $\tau$ are indeed much larger than those involving muons so that the physics discussion of $R_{D^{(\ast )}}$ remains unchanged after setting the couplings to muons to zero.} We find that the allowed  $1\,\sigma$ range in this case reads

\begin{equation}
\label{eq:RDfit}
g_{V_L} \Big{\vert}_{b\to c \tau \nu_{\tau}} \in (0.09,0.13)\,.
\end{equation}
Note that other solutions besides $g_{V_L} >0$ are possible and have been considered in the literature (see e.g. Ref.~\cite{Freytsis:2015qca}).

%\section{Leptoquark models for $R_{K^{(\ast)}}$ and/or $R_{D^{(\ast)}}$}

\section{Which leptoquark model?}

\label{sec:lq-list}

In this Section we list the results for the LQ models proposed to accommodate the $B$-physics anomalies by introducing a single mediator. We adopt the notation of Ref.~\cite{Dorsner:2016wpm} and specify the LQ by their SM quantum numbers, $(SU(3)_c,SU(2)_L)_Y$, where the electric charge, $Q=Y+T_3$, is the sum of the hypercharge ($Y$) and the third weak isospin component ($T_3$). In the left-handed doublets, $Q_i=[(V^\dagger u_L)_i~d_{L\,i}]^T$ and $L_i=[(U\nu_L)_i~\ell_{L\,i}]^T$, the matrices $V$ and $U$ are the CKM and PMNS matrices, respectively. Since the neutrino masses are can be safely ignored for the purposes of this work, we set $U= \mathbb{1}$.

To asses the phenomenological viability of these LQ models, we confront them with the (tree--level) constraints listed in Table~3 of Ref.~\cite{Angelescu:2018tyl} and with constraints arising at loop--level, such as $\tau \to \mu\gamma$, $B_s-\bar{B}_s$ mixing and LFU tests in $Z$ and $\tau$ decays (and $b\to s\nu\bar{\nu}$, which appears only at loop level for $U_1$). However, in the case of vector LQs, we do not consider the constraints induced by the loop effects since they are sensitive to the details of the unknown UV completions, which leads to model dependence (see Ref.~\cite{Angelescu:2018tyl} for a more detailed discussion).

%%%%%%%%%%%%%%%%%%%%%
\begin{table}[t!]

\caption{ \sl \small Summary of the LQ models which can accommodate $R_{K^{(\ast)}}$ (first column), $R_{D^{(\ast)}}$ (second column), and both $R_{K^{(\ast)}}$ and $R_{D^{(\ast)}}$ (third column) without inducing other phenomenological problems. The \textcolor{red}{\xmark}$^{\textcolor{blue}{\ast}}$ symbol means that the discrepancy can be alleviated, but not fully accommodated.}

\renewcommand{\arraystretch}{1.6}
\centering
\begin{tabular}{|c|c|c||c|}
\hline 
Model & $R_{K^{(\ast)}}$ & $R_{D^{(\ast)}}$ & $R_{K^{(\ast)}}$ $\&$ $R_{D^{(\ast)}}$\\ \hline\hline
$S_1$	& \,\,\color{red}\xmark$^{\color{blue}\ast}$ & $\color{blue}\checkmark$	& \,\,\color{red}\xmark$^{\color{blue}\ast}$	\\
$R_2$	&	\,\,\color{red}\xmark$^{\color{blue}\ast}$ & $\color{blue}\checkmark$	&\color{red}\xmark\\
$\widetilde{R_2}$	& \color{red}\xmark	&\color{red}\xmark	 &\color{red}\xmark	\\ 
$S_3$	& $\color{blue}\checkmark$	&\color{red}\xmark	 &\color{red}\xmark	\\   \hline
$U_1$	& $\color{blue}\checkmark$	& $\color{blue}\checkmark$	& $\color{blue}\checkmark$	\\
$U_3$	& $\color{blue}\checkmark$	&\color{red}\xmark	&\color{red}\xmark	\\  \hline
\end{tabular}
\label{tab:LQ-lists} 
\end{table}

Our findings are summarized in Table~\ref{tab:LQ-lists}, from which we learn that $U_1$ is the only single LQ model that can simultaneously  accommodate $R_{K^{(\ast)}}$ and $R_{D^{(\ast)}}$, in agreement with the findings of Ref.~\cite{Buttazzo:2017ixm}. Note that our conclusions can also serve as a guideline for future studies if one of the anomalies disappear.

\section{Revisiting $U_1=(\mathbf{{3}},\mathbf{1})_{2/3}$}
\label{sec:results}

\subsection{Constraints}

As pointed out in the previous section, the only LQ that can provide a simultaneous explanation to the anomalies in $b\to s$ and $b\to c$ transitions is $U_1=(\mathbf{\bar{3}},\mathbf{1})_{2/3}$,~\cite{Buttazzo:2017ixm} the weak singlet vector LQ. In this Section we briefly discuss the phenomenological status of the $U_1$ scenario. To constrain the model parameters of this scenario, we use the low-energy physics observables which receive the tree-level contributions from the $U_1$ exchange. We also compare these results with the ones deduced from the experimental bounds based on direct searches at the LHC. We refer the reader to Ref.~\cite{Angelescu:2018tyl} for a more detailed discussion of all these constraints.

The most genera $U_1$ Lagrangian consistent with the SM gauge symmetry allows couplings to both left-handed and right-handed fermions:

\begin{equation}
\label{eq:lag-U1}
\mathcal{L}_{U_1} = x_L^{ij} \, \bar{Q}_i \gamma_\mu U_1^\mu L_j + x_R^{ij} \, \bar{d}_{R\,i} \gamma_\mu  U_1^\mu \ell_{R\,j}+\mathrm{h.c.},
\end{equation}
where $x_L^{ij}$ and $x_R^{ij}$ are the couplings. To satisfy both $R_{K^{(\ast)}} < R_{K^{(\ast)}}^\mathrm{SM}$ and $R_{D^{(\ast)}}> R_{D^{(\ast)}}^\mathrm{SM}$, we assume the following structure for the coupling matrices:
%%%%%%%%%%%%%%%%%%%%%%
\begin{equation}
\label{eq:yL-U1}
x_L  = \begin{pmatrix}
0 & 0 & 0\\ 
0 & x_L^{s\mu} & x_L^{s\tau}\\ 
0 & x_L^{b\mu} & x_L^{b\tau}
\end{pmatrix}\,, \qquad\qquad x_R = 0\,,
\end{equation}
%%%%%%%%%%%%%%%%%%%%%%
where the left--handed couplings to the first generation are set to zero in order to avoid conflict with experimental limits on $\mu-e$ conversion on nuclei, atomic parity violation and on $\mathcal{B}(K\to \pi \nu\bar{\nu})$. The right--handed couplings are set to zero because otherwise they would generate Wilson coefficients are disfavored by the current $b\to s$ data. In the following, we will call the scenario defined by Eq.~\eqref{eq:yL-U1} the minimal $U_1$ model.

Under these assumptions, the contribution to the effective Lagrangian~\eqref{eq:hamiltonian-bsll} amounts to

\begin{equation}
\label{eq:C9-U1}
 C_{9} = - C_{10} = -\dfrac{\pi v^2}{ V_{tb}V_{ts}^\ast \alpha_{\mathrm{em}}} \dfrac{x_L^{s \mu} \left(x_L^{b \mu}\right)^\ast}{m_{U_1}^2}\,,
\end{equation}

\noindent as required by the observation of $R_{K^{(\ast)}}^{\mathrm{exp}}<R_{K^{(\ast)}}^{\mathrm{SM}}$. Furthermore, this scenario also contributes to $b\to c\tau \bar{\nu}_{\tau}$ by giving rise to the effective coefficient
\begin{align}
\label{eq:gV-U1}
g_{V_L} = \dfrac{v^2\,\left(V x_L\right)_{c\tau}\left(x_L^{b\tau}\right)^\ast}{ 2 V_{cb}\, m_{U_1}^2 } =  \dfrac{v^2}{2 m_{U_1}^2} \left( x_{L}^{b\tau} \right)^\ast \left[ x_L^{b\tau} + \dfrac{V_{cs}}{V_{cb}} x_L^{s\tau} \right] \,,
\end{align}
whose leading term implies $g_{V_L}>0$, in agreement with the observed $R_{D^{(\ast)}}^{\mathrm{exp}}>R_{D^{(\ast)}}^{\mathrm{SM}}$.

\subsection{Results and predictions}

We now present the results of our analysis. For definiteness, we set $m_{U_1} = 1.5$~TeV, which is the lowest $U_1$ mass not yet excluded by vector LQ pair production searches at the LHC.~\cite{CMS:2018bhq} The resulting parameter space will then be used to show our predictions for two LFV processes, $B \to K \mu \tau$ and $\tau \to \mu \phi$. 

For our analysis, we first find a best fit point by minimizing a $\chi^2$-function built from the flavor observables listed in Table~3 of Ref.~\cite{Angelescu:2018tyl}. We find $\chi^2_{\rm min} = 6.61$ for
\begin{align}
\label{eq:chiSQmin_U1}
x_L^{s\mu} \approx -10^{-2}\,, \qquad x_L^{b\mu} \approx 0.25\,, \qquad x_L^{s\tau} \approx 4.4\times 10^{-3}\,, \qquad x_L^{b\tau} \approx 2.81\,.
\end{align} 

\noindent We then perform a random scan over the values of the four left-handed couplings shown in Eq.~\eqref{eq:yL-U1}, and enforce perturbativity, $|x^{ij}_L| \leq \sqrt{4\pi}$. We select only the points which satisfy $\Delta \chi^2 ({\rm par}) \equiv \chi^2 ({\rm par}) - \chi^2_{\rm min} \leq 6.18 $, i.e. within $2\,\sigma$ from the best fit point. The selected points are then confronted with the limits deduced from the direct LHC searches in $\ell \ell$ ($\ell$ =$\mu$, $\tau$) final states (see the discussion in Ref.~\cite{Angelescu:2018tyl}). In the plots presented in this Section, the points excluded by direct searches based on current LHC data (36~fb$^{-1}$) are shown in grey. The red points are those that, according to our projections, will be excluded at 300~fb$^{-1}$. Finally, the blue points are those that would survive. 

\begin{figure}[h!]
  \centering
\includegraphics[width=0.45\textwidth]{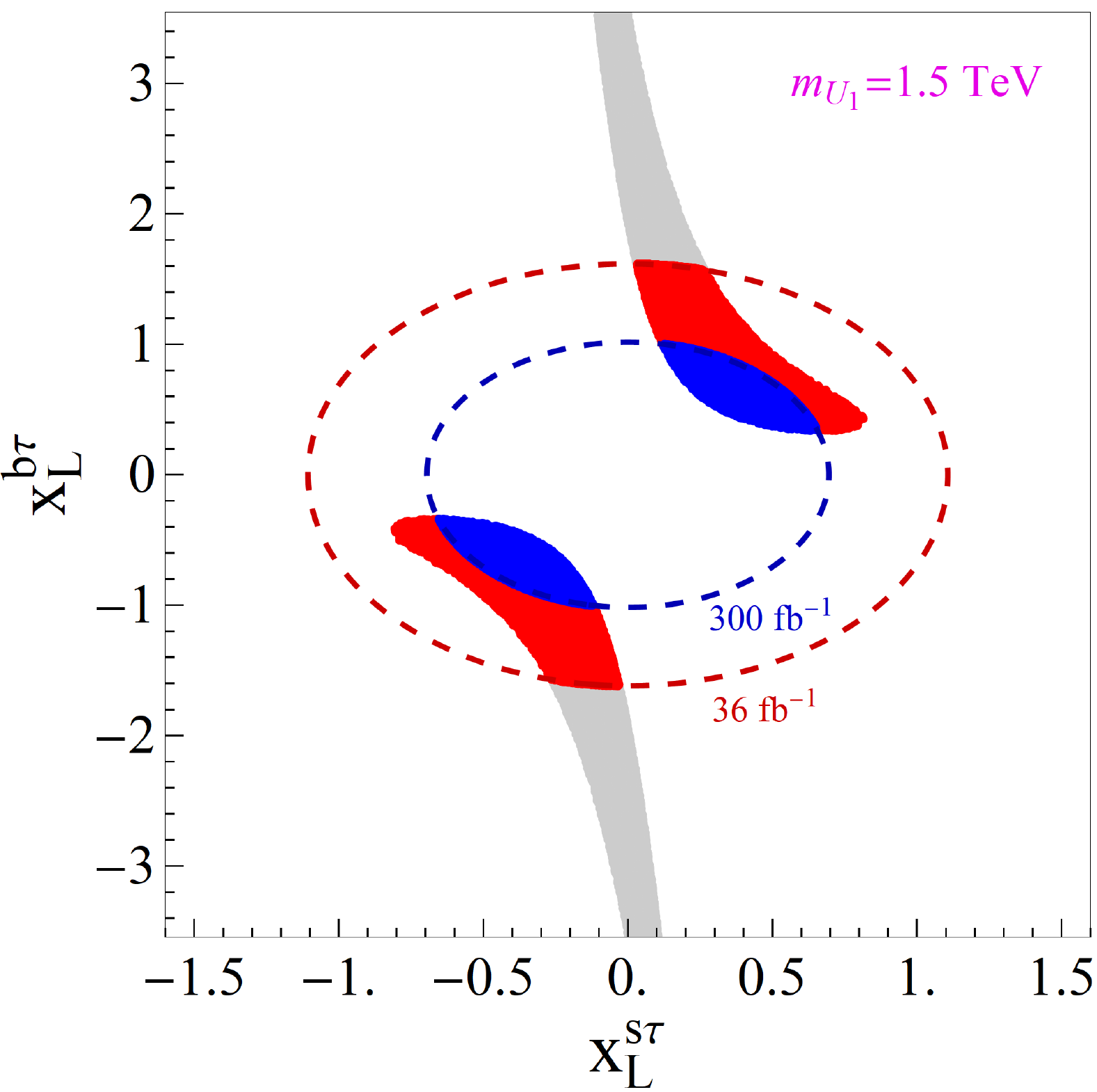} \quad
\includegraphics[width=0.45\textwidth]{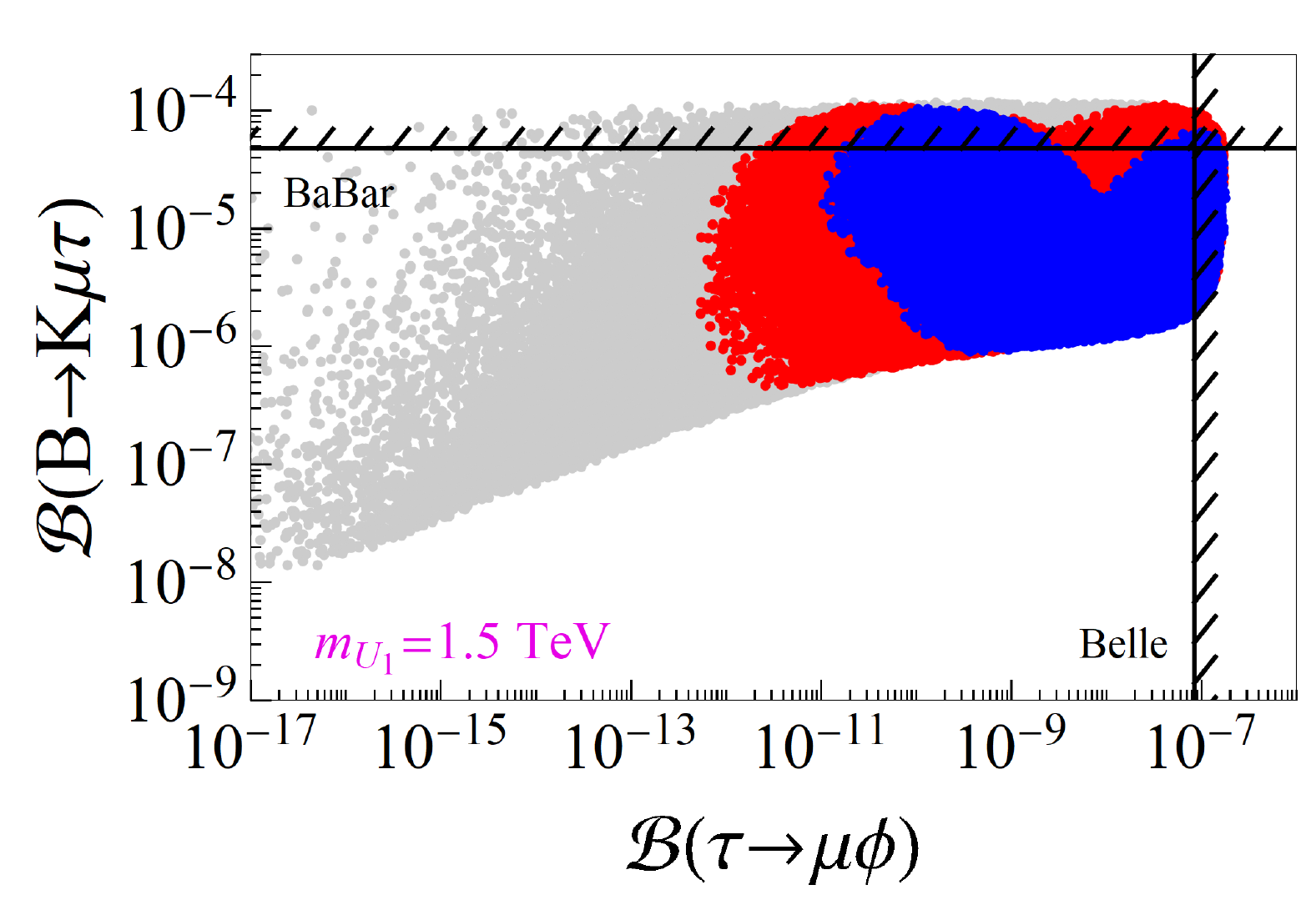}
  \caption{ \sl \small Left panel: the correlation between the couplings $x_L^{s\tau}$ and $x_L^{b\tau}$ allowed by flavor constraints is plotted for $m_{U_1}=1.5$~TeV. Left panel: $\mathcal{B}(B\to K\mu\tau)$ is plotted against $\mathcal{B}(\tau\to\mu\phi)$ for the $U_1$ model. Current bounds on these two decays, as respectively established by BaBar and by Belle are also shown. In both panels, gray points are excluded by current LHC data ($36~\mathrm{fb}^{-1}$) on $pp\to \ell\ell$ ($\ell=\mu,\tau$). The future LHC sensitivity is depicted by the red points, which were obtained by extrapolating current data to $300~\mathrm{fb}^{-1}$. Blue points are allowed by all constraints, including the extrapolated LHC results to $300~\mathrm{fb}^{-1}$.}
  \label{fig:couplings-tau-U1}
\end{figure}

We show, in the left panel of Fig.~\ref{fig:couplings-tau-U1}, the correlation between the two LQ couplings entering Eq.~\eqref{eq:gV-U1} for $m_{U_1}=1.5$~TeV. One observes that the experimental value of $R_{D^{(\star)}}$ forces $x_L^{b\tau}$ to be different from zero, thus bounding its absolute value from below. Similarly, the $R_{K^{(\star)}}$ measured value pushes both $x_L^{b\mu}$ and $x_L^{s\mu}$ to be different from zero (see Eq.~\eqref{eq:C9-U1}). Even though the measurements of low-energy observables allow for $x_L^{s\tau}=0$, we see that current LHC data exclude this possibility, bounding $|x_L^{s\tau}|$ from below. Moreover, our projected bound for 300~fb$^{-1}$ will push the lower limit on $|x_L^{s\tau}|$ even further away from $0$.

%%%%%%%%%%%%%%%%%%
%%%%%%%%%%%%%%%%%%

We then show in the right panel of Fig.~\ref{fig:couplings-tau-U1} our prediction for the correlation of two LFV observables, $\mathcal{B}(\tau \to \mu \phi)$ and $\mathcal{B}(B \to K \mu \tau)$, with the hatched black lines denoting the current experimental bounds on these processes. Again, $m_{U_1}$ is set to 1.5~TeV. The fact that the LHC sets a lower bound on the absolute value of $|x_L^{s\tau}|$ has a dramatic impact on the amount of LFV predicted by the $U_1$ model: the current and future LHC limits lead to a lower bound of $\mathcal{O}(10^{-7})$ for $\mathcal{B}(B \to K \mu \tau)$. We see that improving the experimental bound on $\mathcal{B}(B \to K \mu \tau)$ at LHCb and/or Belle~II can have a major impact on the model building by further restraining the parameter space.

\section{Summary}
\label{sec:conclusions}

In this work we revisited the single LQ solutions to the $B$-physics anomalies, $R_{D^{(\ast)}}^\mathrm{exp}> R_{D^{(\ast)}}^\mathrm{SM}$ and/or $R_{K^{(\ast)}}^\mathrm{exp}< R_{K^{(\ast)}}^\mathrm{SM}$. When confronted with constraints coming from the low-energy flavor physics observables and from direct searches at the LHC, we find that none of the scalar LQs alone, with mass $m_\mathrm{LQ}\simeq 1$~TeV, can provide an NP model capable of simultaneously explaining both anomalies. 

In the case of vector LQs, by focusing only on the tree level observables, we confirm that the weak singlet vector LQ ($U_1$) of mass $m_\mathrm{LQ}\simeq 1\div 2$~TeV can indeed accommodate both $R_{D^{(\ast)}}^\mathrm{exp}> R_{D^{(\ast)}}^\mathrm{SM}$ and $R_{K^{(\ast)}}^\mathrm{exp}< R_{K^{(\ast)}}^\mathrm{SM}$, in its minimal version, i.e. with $x_R = 0$.~\cite{Buttazzo:2017ixm} We observe a pronounced complementarity of the low-energy (flavor physics) constraints with those obtained from direct searches. In particular, assuming perturbative couplings, we find a lower LFV bound of $\mathcal{B}(B\to K\mu\tau) \gtrsim \mathrm{few} \times 10^{-7}$ for $m_{U_1} = 1.5$~TeV (and in fact for any $m_{U_1}$ explaining the anomalies, see Ref.~\cite{Angelescu:2018tyl}). Improving the current experimental bound on $\mathcal{B}(B\to K\mu\tau)$ by two orders of magnitude can therefore either exclude or, if observed, corroborate the validity of the minimal $U_1$ scenario.

\section*{Acknowledgments}

A.A. would like to thank the organizers for the opportunity of presenting this work and for recreating once again the stimulating atmosphere of Moriond. This work is supported by University of Nebraska-Lincoln, National Science Foundation under grant number PHY-1820891, and the NSF Nebraska EPSCoR under grant number OIA-1557417.

\end{document}